
\input phyzzx
\hfuzz 50pt
\font\mybb=msbm10 at 10pt

\def\Bbb#1{\hbox{\mybb#1}}

\def\bR{\Bbb {R}}
\def\bE{\Bbb {E}}

\def\bfomega{\omega\kern-7.0pt \omega}

\magnification=950


\REF\becker{K. Becker, M. Becker and A. Strominger, {\sl Fivebranes, Membranes
and non-perturbative string theory}, Nucl. Phys. {\bf B456} (1995), 130.} 
\REF\ppd {G.W.Gibbons \& G.Papadopoulos,{\sl Calibrations 
and Intersecting Branes}, hep-th/9803163.}
\REF\jpg {J.P.Gauntlett , N.D.Lambert \& P.C.West,{\sl Branes 
and Calibrated Geometries}, hep-th/9803216.}
\REF\jose{J. M. Figueroa-O'Farril, {\sl Intersecting Brane
Geometries}, hep-th/9806040. }
\REF\witten{E. Witten, {\sl Solutions of Four-Dimensional Field Theories
via  M Theory}, Nucl. Phys. {\bf B500} (1997), 3; hep-th/9703166.}
\REF\hvl {R.Harvey \& H.B.Lawson,{\sl Calibrated Geometries}, 
Acta.Math. {\bf{148}} (1982) 47.}
\REF\hv {R.Harvey,{\sl Spinors and Calibrations},Academic
Press (1990),New York.}
\REF\malda{J. Maldacena, {\sl The Large N Limit of Superconformal
Field Theories}, Adv. Theor. Math. Phys. {\bf 2} (1998) 231.}
\REF\rw{R. Rohm and E. Witten, {\sl The Antisymmetric Tensor Field
in Superstring Theory},  Annals Phys. {\bf 170} (1986) 454. }
\REF\gary {G.W. Gibbons, G. T Horowitz and P.K. Townsend, {\sl Higher
Dimensional Resolution of Dilatonic Black Hole Singularities}, Class. Quantum
Grav. {\bf 12} (1995), 297; hep-th/9410073.}
\REF\howe{ P.S. Howe and E. Sezgin, {\sl  D=11, p=5}, Phys. Lett.
{\bf B394} (1997) 62; hep-th/9611008.} 
\REF\claus{P. Claus, R. Kallosh, J. Kumar, P.K. Townsend and A. Van Proeyen,
{\sl Conformal Theory of M-2, D-3, M-5 and `D-1+D-5' Branes}, JHEP 9806 (1998)
(008), hep-th/9801206.}
\REF\bkop {E.Bergshoeff,R.Kallosh,T.Ortin \& G.Papadopoulos,
{\sl Kappa-Symmetry, Supersymmetry and
Intersecting Branes},Nucl.Phys. {\bf B502} (1997) 149.}
\REF\jap{Y. Imamura, {\sl Supersymmetries and BPS 
Configurations on Anti-de Sitter Space}, hep-th/9807179. }
\REF\bilal{A. Bilal and C. S. Chu, {\sl D3 Brane(s) in $AdS_5\times S^5$ and
{\cal N}=4,2,1 SYM}, hep-th/9810195.}
\REF\sonia{S. Stanciu, {\sl D-Branes in Kazama-Susuki Models},
hep-th/9708166.}


\Pubnum{ \vbox{ \hbox{DAMTP-1999-21}\hbox{} } }
\date{February 1999}
\pubtype{}
\titlepage
\title{AdS Calibrations}
\author{J.Gutowski \& G.Papadopoulos}
\address{DAMTP, Silver Street, University of Cambridge,
 Cambridge CB3 9EW}

\pagenumber=1



\def\C{\mkern1mu\raise2.2pt\hbox{$\scriptscriptstyle|$}\mkern-7mu{\rm C}}

\def\pd{\partial_}

\def\m{\mu}
\def\n{\nu}
\def\l{\lambda}
\def\a{\alpha}
\def\b{\beta}

\def\p{\rho}
\def\s{\sigma}
\def\t{\tau}
\def\x{\chi}
\def\g{\gamma}
\def\f{\phi}
\def\d{\delta}
\def\e{\epsilon}

\def\Om{\Omega}
\def\X{\Theta}


\font\mybb=msbm10 at 10pt

\def\Bbb#1{\hbox{\mybb#1}}

\def\bR{\Bbb {R}}
\def\bE{\Bbb {E}}

\def\bfomega{\omega\kern-7.0pt \omega}


\abstract{We present a new bound for the worldvolume actions of branes
with a Wess-Zumino term. For this we introduce a  generalization of 
calibrations for which the calibration form is not closed. We then
apply our construction to find the M-5-brane worldvolume solitons
in an AdS background that saturate this bound. We show that these worldvolume
solitons are  supersymmetric
and that they satisfy
differential equations which generalize those of standard calibrations.}

\endpage


\chapter{Introduction}

The dynamics of branes in a flat background is described, in the
absence of Born-Infeld type  fields, by a Nambu-Goto type of action.
The solutions of the field equations of such an action are minimal 
manifolds. A subclass of solutions of the Nambu-Goto action preserve
a proportion of spacetime supersymmetry and saturate a bound. The most elegant
way to formulate this bound and find the proportion of supersymmetry preserved
by a solution is  in terms of calibrations
 [\becker, \ppd, \jpg, \jose]. Because
brane worldvolume solitons have the bulk interpretation of intersecting branes,
recently there has been much activity in applying calibrations to find 
worldvolume solitons of branes. These  are then used to investigate the quantum
theory of certain gauge theories [\witten].
Despite the many applications that calibrations have in string and M-theory,
there are certain limitations. One is that there does not seem to be a
generalization of the
calibration bound in the presence of Born-Infeld type fields on the brane
worldvolume. Another limitation is that the description of brane dynamics on
curved backgrounds may also require the modification of the Nambu-Goto action
by Wess-Zumino terms. In the presence of such Wess-Zumino terms, the
standard argument used to establish the calibration bounds does not apply.

In this letter, we shall modify the definition of calibrations of [\hvl, \hv]
to establish a new bound for the brane worldvolume actions
with certain types of Wess-Zumino terms. As for 
standard calibrations,
these new calibrations are associated with a calibration k-form $\phi$ which
satisfies the inequality
$$
\phi({\buildrel \rightarrow \over\eta})\leq 1
\eqn\intone
$$
when evaluated at any appropriately normalized k-co-form ${\buildrel 
\rightarrow
\over\eta}$, but in this case 
$\phi$ is not {\it closed}, i.e.
$$
d\phi\not=0\ .
\eqn\inttwo
$$
This modification in the definition of a
calibration allows us to relate  the form field
strength
$F$ associated with the Wess-Zumino term of the brane action 
to the calibration
form as $\tilde F=d\phi$. Let ${\rm Vol}(M)$ and ${\rm WZ}(M)$ be the volume
and the Wess-Zumino term of a manifold $M$, respectively. For  static
worldvolume configurations, we shall establish a bound
$$
{\rm Vol}(M)+{\rm WZ}(M)\leq {\rm Vol}(M')+{\rm WZ}(M')\ ,
\eqn\inthree
$$
where $M$ is a k-dimensional manifold for which $\phi({\buildrel \rightarrow
\over\eta_M})=1$ and ${\buildrel \rightarrow
\over\eta_M}$ is its co-volume.
We shall  apply our method to find the calibrations associated
with the M-5-brane in an $AdS_7\times S^4$ background. Such backgrounds
have been used to formulate a correspondence between
 gauge and string theories
[\malda]. For this, we shall determine the calibration forms and relate them to
the Wess-Zumino term.
We shall then find that many of the differential equations associated
with standard calibrated surfaces in flat spaces admit a natural 
generalization on $AdS_7\times S^4$.

\chapter{Generalized Calibrations}

Let $(N, \tilde g)$ be a n-dimensional oriented Riemannian manifold equipped
with a metric $\tilde g$ and
$G(k,T_pN)$ be the Grassmannian of oriented k-planes in the 
tangent space $T_pN$
at the point $p\in N$.
 Then for $\x \in G(k,T_pN)$ , there exists an 
orthonormal basis $\lbrace e_1 ... e_n \rbrace$ of
$T_pN$ such that $\lbrace e_1 ... e_k \rbrace$ is a 
basis of $\x$. The co-volume of $\x$ is
then 

$$
{\buildrel \rightarrow \over \x} = e_1 \wedge ... \wedge e_k.
\eqn\eqa
$$

We define  a calibration of degree $k$ on an open subset 
$U \subset N$ to be a k-form $\f$ with the
 following properties:

\item{(i)} At each $p \in  U$, 
$$
{\f_p} ( {\buildrel \rightarrow \over \x}) \leq 1
\eqn\twoone
$$
 for all $\x \in 
G(k,T_pN)$, and

\item{(ii)}the 
contact set 
$$
G( \f )= \lbrace \x \in G(k,T_pN):{\f} 
({\buildrel \rightarrow \over \x}) =1 \rbrace 
\eqn\eqb
$$
is non-empty.
Now  a k-dimensional 
submanifold $M$ of $N$ is calibrated if 
$$
\f_p ( {\buildrel \rightarrow \over \eta_M} ) =1
\eqn\eqc
$$
for every $p \in M$, where ${\buildrel \rightarrow \over \eta_M}$ is
the co-volume of $M$. 
 In the conventional definition of a calibration [\hvl,\hv] it is further
 required
that
$$
d\f=0\ .
\eqn\twotwo
$$
To distinguish between the 
calibrations according to the old and new definitions, we shall refer
to the former as \lq\lq standard" calibrations and to the latter as 
\lq\lq generalized" calibrations or simply calibrations. The generalized
calibrations are a special cases of the $\varphi$-geometries of [\hvl].

It remains to establish the bound. For this, let $M$ be a 
calibrated submanifold
of $N$ and $U\subset M$. Then let $V$ be a deformation of $U$ such that $U,V$
have the same boundary
$\partial U=\partial V$. Then we have
$$
{\rm Vol}(U)=\int \phi({\buildrel \rightarrow \over \eta_U}) \mu_U=\int
\phi({\buildrel \rightarrow \over \eta_V}) \mu_V+\int 
d\phi({\buildrel \rightarrow
\over \eta_D})\mu_D\leq {\rm Vol}(V) +\int d\phi({\buildrel \rightarrow
\over \eta_D})\mu_D\ ,
\eqn\wzcali
$$
where $D$ is a disc with boundary $\partial D=U-V$ and $\mu_U$ is the volume
form of $U$ and similarly for the rest. To establish the second equality in
\wzcali,   we have used Stoke's theorem. Next  let $W$ be a reference open set
such that
$\partial W=\partial U=\partial V$,  and  $D_1$ and $D_2$ be
discs such that $\partial D_1= U-W$,  $\partial D_2= V-W$ and $D_2=D_1+D$.
Then the above inequality can be rearranged such that
$$
{\rm Vol}(U)-\int d\phi({\buildrel \rightarrow
\over \eta_{D_1}})\mu_{D_1}\leq {\rm Vol}(V)-\int d\phi({\buildrel \rightarrow
\over \eta_{D_2}})\mu_{D_2}\ .
\eqn\calineq
$$
But using  a slight variation of the definition of a Wess-Zumino term, 
$WZ$, as given in [\rw], we can
set
$$
\eqalign{
WZ(U)&=-\int d\phi({\buildrel \rightarrow
\over \eta_{D_1}})\mu_{D_1}
\cr
WZ(V)&=-\int d\phi({\buildrel \rightarrow
\over \eta_{D_2}})\mu_{D_2}\ .}
\eqn\twothree
$$
with (k+1)-form field strength   $\tilde F=d\phi$.
Substituting the definitions of the Wess-Zumino terms above in \calineq, we
derive the inequality that we have stated in the
 introduction\foot{We do
not consider here the case for which $\tilde F$ 
is closed but not exact. For
our argument, it suffices to assume that 
in the deformation region $D$ of $U$,
 $\tilde
F=d\phi$, where $\phi$ is the calibration form.}. 
This inequality is saturated
by the calibrated submanifolds of
$N$. In the case of a standard calibration $d\phi=0$ and 
so the above inequality
reduces to that of [\hv] for the volumes of $U$ and $V$.

\chapter{Near Horizon M-5-brane Geometries}

Let $\{x^\mu; \mu=0,\dots, 5\}$ be the worldvolume coordinates of M-5-brane.
The  M-5-brane supergravity solution is 
$$
\eqalign{
ds^2&=H^{-{1 \over 3}}\eta_{\mu\nu}dx^\mu dx^\nu+H^{2 \over 3}
\big(dr^2+r^2 ds^2(S^4)\big)
\cr
G_4&=\star dH}
\eqn\eqq
$$
where 
$\eta$ is the Minkowski metric on $\bE^{(1,5)}$, $r$ is a 
radial coordinate and
$$
H=1+{R^3 \over r^3}\ ;
\eqn\eqr
$$
$R$ is a constant. Near the position $r=0$ of the M-5-brane, the solution
becomes
$$
\eqalign{
ds^2&\equiv G_{MN} dx^M dx^N={r \over R}\eta_{\mu\nu}dx^\mu dx^\nu+{R^2 \over
r^2}(dr^2+r^2  ds^2(S^4))
\cr
G_4&=\mu_{S^4} .}
\eqn\eqs
$$
The near horizon geometry of the M-5-brane [\gary] is therefore  $AdS_7
\times S^4$.

The near horizon M-5-brane above  can be used as a background for an
M-5-brane probe with transverse scalars $\{X^a; a=1, \dots 5\}$. The
worldvolume action of the probe which is parallel to the
 background M-5-brane in the static gauge [\howe, \claus] is
$$
S = \int d^{6}x \sqrt{- \det (g_{\m \n})} - \int_D F.
\eqn\exa
$$
where $g_{\m \n}$ is the pull-back to the worldvolume of the
 background $AdS_7 \times S^4$ metric, i.e.
$$
g_{\m \n}={r \over R} \eta_{\m \n}+{R^2 \over r^2}\delta_{ab} \pd{\m}X^a
\pd{\n}X^b\ ,
\eqn\exb
$$
$$
r^2= \delta_{ab} X^a X^b\ ,
\eqn\exc
$$
and
$$
F = d\big({r^3 \over R^3}\big)\wedge dt \wedge dx^1 \wedge dx^2 \wedge dx^3
\wedge dx^4
\wedge dx^5\ .
\eqn\eqt
$$
In the above action we have not included the (self-dual) 3-form worldvolume
field which we have set to zero.

The field equations for the transverse scalars $X$ are
$$
-\pd{\m} ({R^2 \over r^2}\sqrt{-g}g^{\m \n} \pd{\n}X^a)+ \Om X^a=0
\eqn\exi
$$
where 
$$
\Om={3 \over 2r}\sqrt{-g}({1 \over R}g^{\m \n}\eta_{\m \n}-{4 \over r})-
{3r \over R^3}.
\eqn\exj
$$
We shall be considering static  solutions to these field equations which
have a p-brane interpretation, i.e. solutions that are invariant under the
action of the Poincar\'e group on $\bE^{(1,p)}$. For such solution, the
transverse scalars $X$ will depend on $\{x^i; i=1,\dots, 5-p\}$ worldvolume
coordinates. Placing the worldvolume soliton along the first $p$ worldvolume
coordinates of the M-5-brane probe, we find that
$$
\sqrt{- \det (g_{\m \n})}=\big({r \over R}\big)^{p+1\over2} \sqrt{ \det
(g_{ij})}=\sqrt{ \det (\tilde g_{ij})}\ ,
\eqn\exk
$$
where
$$
g_{ij}={r \over R} \delta_{ij}+{R^2 \over r^2}\delta_{ab} \pd{i}X^a
\pd{j}X^b\ ,
\eqn\exbbb
$$
and
$$
\tilde g_{ij}=\big({r \over R}\big)^{p+1\over5-p} g_{ij}\ .
\eqn\exbbb
$$
We remark that the metric $g$ is the pull back of the  metric on  
$H^{6-p}\times
S^4$ with respect to $X$, where
$H^{6-p}$ is a $(6-p)$-dimensional hyperbolic space.  Using all these
redefinitions, the non-trivial part of the field equations for the p-brane
worldvolume soliton can be derived by varying the action 
$$
S_p =\lambda ( \int d^{5-p}x \sqrt{ \det (\tilde g_{ij})} - \int_D \tilde F\ ),
\eqn\exaaa
$$
where
$$
\tilde F=
 d\big({r^3 \over R^3}\big)\wedge dx^1 \wedge \dots
\wedge dx^{5-p}
\eqn\mfs
$$
and $\lambda$ is a constant.  In what follows, we shall apply our
calibrations to find  solutions to field equations of the action
\exaaa. We shall find that that there are generalized analogues
to well known standard K\"ahler, Special Lagrangian (SLAG) and exceptional
calibrations in
$\bR^n$.

The supersymmetry projection operator is

$$
\Gamma = {1 \over 6!\sqrt{-g}}\e^{\m \n \p \s \t \l}
{\hat{\g}}_{\m}{\hat{\g}}_{\n}{\hat{\g}}_{\p}{\hat{\g}}_{\s}
{\hat{\g}}_{\t}{\hat{\g}}_{\l}
\eqn\exd
$$
where ${\hat{\g}}_{\m}$ are the pull-backs of the bulk ${\hat{\Gamma}}$
matrices.
The proportion of supersymmetry preserved by a solution of the field
equations of \exaaa\ is
determined by the  number of linearly independent solutions to  [\bkop]
$$
(1 - \Gamma) \e =0\ ,
\eqn\exh
$$
where $\e$ satisfy the Killing spinor equations for the $AdS_7\times S^4$
background.

\chapter{AdS Hermitian Calibrations}

The AdS Hermitian  calibrations are the analogue to the K\"ahler
 calibrations on
$\bR^n$. To define the degree $2\ell$ AdS Hermitian  calibrations,  we shall 
take
$N$ to be conformal to the space
$H^{2\ell+1}\times S^{2m-1}$ equipped with the metric 
$$
\eqalign{
d\tilde s^2=& \big({r \over R}\big)^{3-\ell\over\ell}
(ds^2(H^{(2\ell+1)})+R^2 ds^2(S^{2m-1}))
\cr
&= 
\big({r \over R}\big)^{3-\ell\over\ell}
\big[{r\over R}ds^2(\bE^{2\ell})+ {R^2\over r^2} ds^2(\bE^{2m})\big]\ ,}
\eqn\threethree
$$
where 
$$
ds^2(H^{(2\ell+1)})= {r\over R}ds^2(\bE^{2\ell})+ {R^2\over r^2} dr^2
\eqn\fourthree
$$ 
is the metric on the $(2\ell+1)$-dimensional
hyperbolic space.
We then use the orthonormal basis 
$$
\eqalign{
e^{ A}&= \big({r \over R}\big)^{3\over2\ell} dx^A\ , \qquad\qquad 
1\leq A\leq 2\ell\ ,
\cr
e^{ A}&=\big({r \over R}\big)^{3-3\ell\over2\ell} dx^A \ , \qquad
2\ell+1 \leq A\leq 2(\ell+m) }
\eqn\fiveone
$$
of the above metric and define an almost complex structure $I$ with  
K\"ahler form
$$
\omega_I=\sum^{\ell+m}_{q=1} e^{ 2q-1}\wedge e^{ 2q}\ .
\eqn\fivetwo
$$
In the coordinate basis $\omega_I$ can be written as
$$
\omega_I=\big({r \over R}\big)^{3\over\ell} \sum^\ell _{i=1}
dx^{2i-1}\wedge dx^{2i}+\big({r
\over R}\big)^{3-3\ell\over\ell} \sum_{a=1}^m 
dx^{2i-1+2\ell}\wedge dx^{2i+2\ell}\ .
\eqn\fivethree
$$
Using this, we observe that  $I$ is a complex structure because it is
a 
constant
tensor  in
the coordinate basis.
We then proceed to define the calibration form
$$
\phi={1\over \ell!} \wedge^{\ell}\omega_I\ .
\eqn\fivefour
$$
The form $\phi$ satisfies  Wirtinger's inequality. This is because
in the orthonormal basis the calibration form in the same as that of the
K\"ahler calibration.  The subspaces of $Gr(2\ell, T_pN)$ that
saturate the
 bound
are those that are complex with respect to the complex structure $I$ defined
above. So $\phi$ defines a calibration in the generalized sense with contact
set
$$
G(\phi)=U(\ell+m)/U(\ell)\times U(m)\ .
\eqn\fivefive
$$
 The calibrated submanifolds are the
complex submanifolds of
$N$ with respect to the {\it complex} structure
$I$.
  
Our AdS  Hermitian  calibrated manifolds correspond to M-5-brane 
worldvolume p-brane solitons for which $p=5-2\ell$ and with 
$2m$ non-vanishing transverse scalars $X$. To establish this, it remains to
show that $\tilde F=d\phi$, where $\tilde F$ is given in \mfs. 
This  can be verified by a straightforward computation. The argument
for computing the supersymmetry preserved by brane worldvolume solitons
associated with standard calibrations of  [\ppd, \jpg] can be easily
 generalized in
this case. In particular, we find that  the above
worldvolume solitons preserve  $1/2^{\ell+m}$ of spacetime supersymmetry.
In an orthonormal frame of the metric $G$, the projection operators are
precisely those that have been found for the standard K\"ahler case.

\section{ AdS  SAS  Calibrations} 

Special Almost Symplectic (SAS) calibrations are the analogues of the standard
Special Lagrangian (SLAG) calibrations.   To define a degree $\ell$ 
AdS SAS calibrations,  we shall take
$N$ to be conformal to the space
$H^{\ell+1}\times S^{\ell-1}$ equipped with metric 
$$
\eqalign{
d\tilde s^2&= \big({r \over R}\big)^{6-\ell\over\ell}
(ds^2(H^{(\ell+1)})+R^2 ds^2(S^{\ell-1}))
\cr &= \big({r \over
R}\big)^{6-\ell\over\ell}
\big[{r\over R}ds^2(\bE^{\ell})+ {R^2\over r^2} ds^2(\bE^{\ell})\big]\ ,}
\eqn\fivesix
$$
where 
$$
ds^2(H^{(\ell+1)})= {r\over R}ds^2(\bE^{\ell})+ {R^2\over r^2} dr^2
\eqn\fiveseven
$$ 
is the metric on the $(\ell+1)$-dimensional
hyperbolic space.
We then use the orthonormal basis 
$$
\eqalign{
e^{ A}&= \big({r \over R}\big)^{3\over\ell} dx^A\ , \qquad  1\leq
A\leq \ell\ ,
\cr
e^{ A}&=\big({r \over R}\big)^{6-3\ell\over2\ell} dx^A \ , \qquad
\ell+1 \leq A\leq 2\ell }
\eqn\fiveeight
$$
of the above metric and define an almost symplectic form 
$$
\omega=\sum^{\ell}_{q=1} e^{ q}\wedge e^{ q+\ell}\ .
\eqn\ninea
$$
Let $I$ be the almost complex structure with K\"ahler form $\omega$.
The SAS calibration form is
$$
\phi={\rm Re} (e^1+ie^{\ell+1})\wedge \dots\wedge (e^{\ell}+ie^{2\ell})
\eqn\nineb
$$
Since  the form $\phi$ coincides in an orthonormal basis with that used as 
a calibration form in the SLAG case, it satisfies the same inequalities as
those that can be found in [\hvl].  These inequalities are saturated by the
special Lagrangian $\ell$-planes in $Gr(\ell, \bR^{2\ell})$.  So
$\phi$
 defines a
calibration in the generalized sense with contact set 
$$
G(\phi)=SU(\ell)/SO(\ell)\ .
\eqn\ninec
$$
 A consequence of this is that
$$
{\rm Im} (e^1+ie^{\ell+1})\wedge \dots\wedge (e^{\ell}+ie^{2\ell})=0\ ,
\eqn\imag
$$
and conversely, any plane that satisfies \imag\ is special Lagrangian. 
 
To derive the equations that our calibrated manifolds satisfy,  we first
rewrite the almost symplectic form in a coordinate basis as
$$
\omega=\big({r \over R}\big)^{12-3\ell\over2\ell} 
\sum^\ell_{i=1} dx^i\wedge dy_i
\eqn\nind
$$
where $y_i=x^{i+\ell}$.  We expect that the calibrated manifolds $M$ can be
locally expressed as $y_i=y_i(x^j)$. Moreover, the pull-back of $\omega$ on $M$
must vanish. This implies that there is a function $f=f(x^i)$ such that
$$
y_i=\partial_i f\ .
\eqn\tena
$$
Substituting this into \imag, we find that the function $f$ satisfies
$$
{\rm Im}\big[ {\rm det}\big (\delta_{ij}+i{R\over r}^{3\over2}
\partial_i\partial_j f)\big]=0\ ,
\eqn\sas
$$
where
$$
r^2= \delta^{ij} \partial_i f \partial_j f\ .
\eqn\tenn
$$

Degree two SAS calibrations are the same as the degree two AdS Hermitian 
calibrations investigated in the previous section. For degree three SAS
calibrations the equation \sas\ reduces to
$$
{(\delta^{mn} \partial_mf \partial_n f)^{3\over2}\over R^3}\,
\delta^{ij}\partial_i\partial_j f= {\rm det}(\partial_i\partial_j f)\ ,
\eqn\threesas
$$
where $i,j,m,n=1,2,3$. This equation corresponds to the Monge-Amp\'ere
equation of the degree three SLAG calibrations.
To find solutions to this equation, we assume that $f$ is spherically
symmetric, i.e. $f=f(u)$ where $u^2=\delta_{ij} x^i
x^j$. Then the equation \threesas\ becomes
$$
({u^2f' \over R^3}-1)f''+{2u \over R^3}{f'}^2=0
\eqn\sli
$$
where $f'={d\over du}f$. A solution is
$$
f(u)=-R^3 \int_{u_0}^u {1 \over v^2}L(\b v^2) dv\ ,
\eqn\slj
$$
where $u_0$,$\b$ are constants and $L(v)$ is the principal
 branch of the Lambert
function which is analytic at $v=0$ and satisfies
$$
L(v)e^{L(v)}=v\ .
\eqn\slk
$$
For degree four SAS calibrations, the equation \sas\ reduces to
$$
{(\delta^{mn} \partial_mf \partial_n f)^{3\over2} \over R^3}
\d^{ij}\pd{i}\pd{j}f=\sum_{k} Det_{k|k}(\pd{i}\pd{j}f)
\eqn\slp
$$
where $i,j, m,n=1,2,3,4$ and $Det_{k|k}(\pd{i}\pd{j}f)$ is the determinant
of the matrix $(\pd{i}\pd{j}f)$ with the $k$-column and $k$-row missing. A
spherically symmetric solution is
 $$
f(u)={5 R^3 \over u}\ ,
\eqn\sls
$$
where where $u^2=\delta_{ij} x^i x^j$.

Our AdS SAS   calibrated manifolds correspond to p-brane worldvolume
M-5-brane solitons for which $p=5-\ell$ and with 
$\ell$ non-vanishing transverse scalars $X$. To establish this, it remains to
show that $\tilde F=d\phi$ for $\ell=5-p$, where $\tilde F$ is given in \mfs. 
This  can be verified by a straightforward computation for the cases $\ell=3,4$
but it is not the case for $\ell=5$. The argument for computing the 
supersymmetry
preserved by brane worldvolume solitons associated with standard calibrations
of  [\ppd, \jpg] can be easily generalized in this case. In
particular,
 we find that 
the SAS worldvolume M-5-brane solitons preserve  $1/2^{\ell}$ of spacetime
supersymmetry. In the orthonormal frame of the metric $G$, the projection
operators are precisely those that have been found for the standard SLAG case.

\chapter{AdS Exceptional  Calibrations}

The construction of AdS exceptional  calibrations proceeds in the same way
as for Hermitian and for SAS calibrations. One begins with the
 standard constant
exceptional calibration form in each case and constructs the calibration
form of the AdS calibration by contracting it with an orthonormal basis
of the associated curved metric. It is clear that all the necessary
inequalities follow  by those presented in [\hvl] for the exceptional cases.
The contact sets of the AdS exceptional  calibrations are those of the standard
exceptional ones. In all these cases
$d\phi=\tilde F$ and so the calibrated manifolds have a p-brane worldvolume
soliton interpretation. The supersymmetry preserved by each such
 solution is the
same as that for p-brane worldvolume
solitons associated with the standard exceptional calibrations.
In what follows we shall simply state the metric on the manifold $N$ and the
equations which the calibrated manifolds satisfy.

\section{  AdS Cayley Calibrations}  

The AdS Cayley  calibration is a degree four calibration in $N$ equipped with
metric
$$
d\tilde s^2={r^{3 \over 2} \over R^{3 \over 2}}\delta_{ij} dx^i dx^j+
{R^{3 \over
2}
\over r^{3
\over 2}}\delta_{ij} dy^i dy^j\ ,
\eqn\ega
$$
where $r^2=\delta_{ij} y^i y^j$ and $i,j=1,\dots, 4$. 
The calibration four-form is constructed from the ${\rm Spin}(7)$ invariant
self-dual form on $\bR^8$. The calibrated manifolds can be locally expressed
as $y^i=y^i(x)$ and $y^i$ satisfy the modified Cayley equations as follows:
$$
\eqalign{
\big({r \over R}\big)^3( \pd{1} \X-\pd{2}\X i-\pd{3}\X j-\pd{4}\X k)=\pd{2}\X
\times 
\pd{3}\X \times \pd{4}\X +  \pd{1}\X \times \pd{3}\X \times \pd{4}\X i
\cr
-\pd{1}\X \times \pd{2}\X \times \pd{4}\X j+\pd{1}\X \times \pd{2}\X \times 
\pd{3}\X k}
\eqn\egc
$$
and
$$
\eqalign{
{\rm Im}[(\pd{1} \X \times \pd{2} \X - \pd{3} \X \times \pd{4} \X )i+(\pd{1} 
\X \times \pd{3} \X + \pd{2} \X \times \pd{4} \X )j
\cr
+(\pd{1} \X \times \pd{4} \X - \pd{2} \X \times \pd{3} \X )k]=0}
\eqn\egd
$$
where $\X=y^1+iy^2+jy^3+ky^4$, $a \times b \times c ={1 \over
2}(a{\bar{b}}c-c{\bar{b}}a)$,
 $a \times b = -{1 \over 2}({\bar{a}}b-{\bar{b}}a)$ and $i,j,k$ are the
imaginary unit quaternions.

\section{ AdS Associative  Calibrations} 
The AdS associative  calibration is a degree three calibration in $N$ 
with metric
$$
d\tilde s^2={r^2 \over R^2}\delta_{ij} dx^i dx^j+{R \over r}\delta_{ab} dy^a
dy^b\ ,
\eqn\egf
$$
where $r^2=\delta_{ab} y^a y^b$, $a,b=1,\dots, 4$ and $i,j=1,\dots,3$.
The  calibration form can be constructed
from the $G_2$ invariant three-form on $\bR^7$ in the way that we have
explained
above. The calibrated manifolds can be locally expressed
as $y^a=y^a(x)$ and $y^a$ satisfy the following equations:
$$
-\big({r \over R}\big)^3(\pd{1} \X i + \pd{2} \X j + \pd{3} \X k) = \pd{1} \X 
\times \pd{2} \X \times \pd{3} \X
\eqn\egh
$$
where $\X=y^1+iy^2+jy^3+ky^4$ and $\times$ are defined  as for the Cayley
case above.

\section {AdS Coassociative   Calibrations}

The AdS coassociative  calibration is a degree four calibration in $N$ equipped
with metric
$$
d\tilde s^2={r^{3 \over 2} \over R^{3 \over 2}}\delta_{ij} dx^i dx^j
+{R^{3 \over
2}
\over r^{3
\over 2}}\delta_{ab} dy^a dy^b\ ,
\eqn\egj
$$
where $r^2=\delta_{ab} y^a y^b$, $a,b=1,\dots, 3$ and $i,j=1,\dots,4$.
The  calibration form can be constructed
from the $G_2$ invariant four-form on $\bR^7$ in the way that we have explained
above. The calibrated manifolds can be locally expressed
as $y^a=y^a(x)$ and $y^a$ satisfy the following equations:
$$
-\big({r \over R}\big)^3(\partial X i + \partial Y j + \partial Z k)
 = \partial X
\times \partial Y \times \partial Z
\eqn\egl
$$
where $X=y^1$, $Y=y^2$, $Z=y^3$ and $\partial =\pd{1}+i\pd{2}+j\pd{3}+k\pd{4}$.
A well-known example of a standard co-associative calibrated manifold is the
Lawson-Osserman co-associative cone. We may generalize this
solution by considering the ansatz
$$
 Xi+Yj+Zk ={\rm Im}( f(u) {\bar{x}} i x )\ , 
\eqn\laws
$$
where $x=x^1+ix^2+jx^3+kx^4$ and
$u^2=\delta_{ij} x^i x^j$. This  solves the equation \egl\  for
$$ 
f(u)=-{4 R^3 \over u^4}+{R^3 e^\a \over 2 u^6}(e^\a \pm
\sqrt{e^\a - 16 u^2})\ ,
 \eqn\modlws
$$
where $\alpha$ is a constant.

\chapter{Conclusions}

We have found that the investigation of 
M-5-brane supersymmetric worldvolume
solitons  on $AdS_7\times S^4$ background 
requires the generalization
of calibrations due to the presence of 
Wess-Zumino terms in the M-5-brane
worldvolume action. We have presented 
such a generalization of
calibrations and we have established a bound that  is saturated by
the calibrated manifolds. Our construction is general 
but when it is applied
to the case of M-5-brane, the functional that  
is minimized is related
to the worldvolume action without Born-Infeld 
type fields. We have also found
that there is a correspondence between 
calibrations on $AdS_7\times S^4$ and
calibrations on
$\bR^{(1,10)}$ which induces  a correspondence
 between the equations
satisfied by  these calibrations.

There are many other cases to consider. For example solutions
to  brane probe actions
in the near horizon geometries of the M-2-brane 
and the D-3-brane; some of
the AdS Hermitian calibrations for the D-3-brane 
have been investigated in
[\jap, \bilal]. Our construction can be generalized to 
treat both these cases. Since brane probes in  generic string and
 M-theory backgrounds which
have  non-vanishing form
field strengths couple with terms that include the Wess-Zumino ones, the
investigation of supersymmetric  worldvolume solutions may
require a further generalization of 
calibrations. One such example
is the case of D-brane probes in 
NS$\otimes$NS backgrounds. Such backgrounds have 
been investigated from the
conformal field theory point of view in [\sonia]. These may
have  applications in
strings, M-theory and differential geometry.

\vskip 1cm
\noindent{\bf Acknowledgments:}   J.G.  thanks EPSRC
 for a studentship.  G.P. is supported by a
University Research Fellowship from the Royal Society.

\refout
\end